\documentstyle[epsf,emulateapj]{article}
\def\ergps{erg s$^{-1}$ }

\def\hmtwo{h$_{\rm 50}^{-2}$ }

\begin{document}
\title{The WARPS survey: III. The discovery of an X-ray luminous galaxy cluster
       at $z=0.833$ and the impact of X-ray substructure on cluster abundance
       measurements}
\author{H.\ Ebeling\altaffilmark{1,9}, L.R.\ Jones\altaffilmark{2,9},
        E.\ Perlman\altaffilmark{3,4}, C.\ Scharf\altaffilmark{3,5,6},
        D.\ Horner\altaffilmark{5,6}, G.\ Wegner\altaffilmark{7},
        M.\ Malkan\altaffilmark{8}, B.\ Fairley\altaffilmark{2}, 
        C.R.\ Mullis\altaffilmark{1}}
\altaffiltext{1}{Institute for Astronomy, 2680 Woodlawn Drive, Honolulu,
       Hawaii 96822, USA}
\altaffiltext{2}{School of Physics and Space Research, University of
       Birmingham, Brimingham B15\,2TT, UK}
\altaffiltext{3}{Space Telescope Science Institute, Baltimore, MD 21218, USA}
\altaffiltext{4}{Department of Physics and Astronomy, Johns Hopkins
	University, 3400 North Charles Street, Baltimore, MD 21218, USA}
\altaffiltext{5}{Laboratory for High Energy Astrophysics, Code 660, NASA/GSFC, 
       Greenbelt, MD 20771, USA}
\altaffiltext{6}{University of Maryland, College Park, MD 20742-2421, USA}
\altaffiltext{7}{Dept. of Physics \& Astronomy, Dartmouth College, 
       6127 Wilder Lab., Hanover, NH 03755, USA}
\altaffiltext{8}{Dept. of Astronomy, UCLA, Los Angeles, CA 90024, USA}
\altaffiltext{9}{Visiting Astronomer at the W.M. Keck Observatory, jointly
                 operated by the California Institute of Technology and the
                 University of California.}

\slugcomment{accepted for publication in ApJ}

\begin{abstract}

The WARPS (Wide Angle ROSAT Pointed Survey) team reviews the
properties and history of discovery of ClJ0152.7--1357, an X-ray
luminous, rich cluster of galaxies at a redshift of $z=0.833$. At
$L_{\rm X} = 8\times 10^{44}$ \hmtwo \ergps ($0.5-2.0$ keV)
ClJ0152.7--1357 is the most X-ray luminous cluster known at redshifts
$z>0.55$. The high X-ray luminosity of the system suggests that
massive clusters may begin to form at redshifts considerably greater
than unity. This scenario is supported by the high degree of optical
and X-ray substructure in ClJ0152.7--1357, which is similarly complex
as that of other X-ray selected clusters at comparable redshift and
consistent with the hypothesized picture of cluster formation by mass
infall along large-scale filaments.

X-ray emission from ClJ0152.7--1357 was detected already in 1980 with
the EINSTEIN IPC. However, because the complex morphology of the
emission caused its significance to be underestimated, the
corresponding source was not included in the cluster sample of the
EINSTEIN Extended Medium Sensitivity Survey (EMSS) and hence not
previously identified. Simulations of the EMSS source detection and
selection procedure performed by us suggest a general, mild bias of
the EMSS cluster sample against X-ray luminous clusters with
pronounced substructure.

If highly unrelaxed, merging clusters are common at intermediate to
high redshift (as is suggested by the current data) they could create
a bias in some samples as the morphological complexity of mergers may
cause them to fall below the flux limit of surveys that make the
implicit or explicit assumption of a unimodal spatial source
geometry. Conversely, the enhanced X-ray luminosity of mergers might
cause them to, temporarily, rise above the flux limit. Either effect
could lead to erroneous conclusions about the evolution of the
comoving cluster space density.  A high fraction of morphologically
complex clusters at high redshift would also call into question the
validity of evolutionary studies (and, specifically, cosmological
conclusions) which implicitly or explicitly assume that the systems
under investigation are virialized. 

\end{abstract}

\keywords{galaxies: clusters: general --- galaxies: clusters: individual
          (ClJ0152.7--1357, MS1054.4--0321, MS1137.5+6625, RX\,J1716.6+6708)
          --- cosmology: observations --- X-rays: general}

\section{Introduction} 

The space density of distant clusters of galaxies is a measurable
quantity whose theoretical value is highly sensitive to the physical
and cosmological parameters of models of structure formation and
evolution (e.g., Oukbir \& Blanchard, 1992; Bahcall \& Cen, 1992;
Viana \& Liddle, 1996; Carlberg et al., 1997; Oukbir \& Blanchard,
1997; Eke et al., 1998).

A large number of independent measurements of the cluster X-ray
luminosity function (XLF) have been performed in the past decade.
Given the diversity of the original observations used in these studies
and of the data analysis techniques applied, the good agreement of
the results is impressive. Virtually all studies agree that the
abundance of clusters of low to intermediate X-ray luminosity ($L_{\rm
X} < 4 \times 10^{44}$ \ergps, $0.5-2.0$ keV) does not change
significantly out to $z\sim 0.8$ (Gioia et al., 1990b; Henry et al.,
1992; Burke et al., 1997; Ebeling et al., 1997; Vikhlinin et al.,
1998a; Jones et al., 1998; Rosati et al., 1998; DeGrandi et al.\ 1999;
Nichol et al., 1999).

At higher X-ray luminosities, however, a consistent picture has yet to
emerge. Reports of strong negative evolution at $L_{\rm X} > 3 \times
10^{44}$ \ergps ($0.3-3.5$ keV) already at moderate redshifts just
beyond $z= 0.3$ (Gioia et al., 1990b; Henry et al., 1992; see also
Nichol et al., 1997, for a contrary result) are supported by the
findings of Vikhlinin et al.\ (1998a), although the latter rest on a
statistically less secure basis.  If these results are correct, a much
more pronounced dearth of X-ray luminous clusters is expected at yet
higher redshift, unless cluster evolution is a strongly non-linear,
almost discontinuous function of X-ray luminosity and redshift.
However, Luppino \& Gioia (1995) show that the cluster XLF at $0.5\la
z \la 1$ is consistent with the one found in the EMSS for the redshift
range $0.3<z<0.6$ (median $z=0.33$), i.e., there appears to be no
further significant evolution of luminous clusters beyond $z\approx
0.6$ (although such evolution is not ruled out).

With only a handful of X-ray luminous clusters currently known at
$z>0.5$, the key to understanding these apparently conflicting results
lies clearly in new discoveries, and more detailed observations, of
X-ray luminous (and, by inference, massive) clusters at high redshift.
Any additional detection of a massive cluster at high redshift (and
certainly at $z\ga 0.8$) is thus of paramount importance as it brings
us one step closer to an accurate measurement of the cluster abundance
at very high redshift, where its sensitivity to evolutionary effects
is greatest.

\section{The significance of distant massive clusters}

With the exception of the Bright SHARC Survey (Nichol et al., 1999),
all of the present deep PSPC cluster surveys\footnote{i.e., the ROSAT
North Ecliptic Pole Survey (Mullis, Gioia \& Henry 1998), the ROSAT
Deep Cluster Survey (Rosati et al., 1995; Rosati et al., 1998), the
southern Serendipitous High-Redshift Archival Cluster Survey (SHARC-S,
Collins et al., 1997; Burke et al., 1997), the Wide Angle ROSAT
Pointed Survey (Scharf et al., 1997, Paper I; Jones et al., 1998,
Paper II; Fairley et al., 1999) and the CfA cluster survey (Vikhlinin
et al., 1998a,b, 1999)} provide sufficient depth to detect a cluster
of $L_{\rm X} > 10^{45}$ \ergps ($0.3-3.5$ keV)\footnote{note that we
use $h=0.5, q_0=0.5$ throughout} out to $z\approx 1$ and beyond. When
the cumulative high-redshift EMSS XLF (median $z=0.33$) is scaled to
the comoving volume corresponding to the redshift range $0.8<z<1$
(i.e., assuming no evolution between $z\approx 0.33$ and $z=0.8-1$),
it predicts about 15 clusters with $L_{\rm X} > 10^{45}$ \ergps
($0.3-3.5$ keV) per steradian, i.e. $4.6\times 10^{-3}$ per square
degree.  Since the mentioned cluster surveys cover only from 18 square
degrees (SHARC-S) to 160 square degrees (CfA cluster survey), the
detection of only very few X-ray luminous clusters in any of these
surveys places significant constraints on the evolution of clusters
and large scale structure in general. Note that this is true only for
the most luminous systems: if the X-ray luminosity criterion is
relaxed and clusters down to $L_{\rm X} = 4\times 10^{44}$ \ergps
($0.3-3.5$ keV) are considered, the expected cluster density in the
same redshift range rises by almost an order of magnitude and any
individual cluster detection becomes much less significant.

We emphasize that, although any detection (X-ray, optical, infra-red)
of massive clusters at very high redshifts is an important discovery
in its own right, it is clusters detected in the course of
statistically complete surveys that bear the most weight. Only the
latter allow the space density of such systems to be quantified and
compared to predictions from theoretical models. Any systematic
effects in the data analysis and interpretation that could cause such
clusters to be missed or misidentified need to be thoroughly
understood and corrected for before conclusions about the physical or
cosmological parameters governing cluster evolution are drawn from
derived statistics such as the cluster XLF.

In the rest of this paper we summarize the current observational
status (Section~\ref{sofar}) and give a short overview of the WARPS
serendipitous cluster survey (Section~\ref{warps}). We then describe
the WARPS discovery of ClJ0152.7--1357, a very X-ray luminous cluster
at $z=0.8325$, and discuss and summarize the results from all
available X-ray observations of this system (Section~\ref{disc_obs}).
Prompted by our finding that ClJ0152.7--1357 was missed in the EMSS,
we take a closer look at how deviations from spherical symmetry in a
cluster's X-ray emission may affect the EMSS cluster sample (and
possible other cluster samples) as a whole (Section~\ref{emss}).
Trying to assess the importance of biases caused by complex cluster
morphology we investigate the prevalence of substructure in distant
clusters (Section~\ref{morph}) before, finally, discussing the
implications of our findings for attempts to constrain cosmological
parameters using X-ray flux limited cluster samples
(Section~\ref{summary}).

\section{Previously known very distant, X-ray luminous clusters of galaxies}
\label{sofar}

Very few clusters of galaxies have been detected at redshifts greater
than 0.8, and even fewer can be called X-ray luminous. Prior to the
discovery of ClJ0152.7--1357, only two X-ray selected clusters were
known at $z>0.8$: MS1054.4--0321 ($z=0.829$, $L_{\rm X} = 1.42 \times
10^{45}$ \ergps in the $0.3-3.5$ keV band\footnote{The given
luminosity was derived from the archival ROSAT HRI observation of this
cluster.}, Donahue et al., 1998) and, much less X-ray luminous,
RX\,J1716.6+6708 ($z=0.813$, $L_{\rm X} = 3.2 \times 10^{44}$ \ergps
in the $0.5-2.0$ keV band, Henry et al.\ 1997, Gioia et al.\
1999). Slightly closer than $z=0.8$ ($z=0.782$, Gioia \& Luppino
1994), but distant and X-ray luminous enough to be noteworthy in this
context, is MS1137.5+6625 ($L_{\rm X} = 1.03 \times 10^{45}$ \ergps in
the $0.3-3.5$ keV band\addtocounter{footnote}{-1}\footnotemark). The
discovery of an even more distant X-ray emitting cluster at $z=1.27$
was recently reported by Rosati et al.\ (1999); however, at $L_{\rm X}
\sim 1.5 \times 10^{44}$ \ergps ($0.5-2.0$ keV) this system is even
less X-ray luminous than RX\,J1716.6+6708.

All other presently known clusters at very high redshift have been
optically selected as projected galaxy overdensities in deep CCD
images (Gunn, Hoessel \& Oke, 1986, GHO; Postman et al., 1996, PDCS,
Scodeggio et al.\ 1999) or were originally detected at radio or
infrared wavelengths (e.g.\ Crawford \& Fabian 1996, Deltorn et al.,
1997, Stanford et al., 1997). Although there are now an impressive
number of optically selected clusters at $z>0.8$ (Postman and
coworkers alone list a dozen clusters at $z\ge 1$ in their PDCS
sample), it ought to be emphasized that, for the majority of these
possibly very distant optical clusters, the published `redshifts' are
estimated statistically and are not the result of actual spectroscopic
measurements. The physical reality of many of these systems thus
remains to be confirmed through either X-ray or extensive
spectroscopic observations. The difficulties inherent to the optical
approach are evidenced by, e.g., the work of Oke, Postman \& Lubin
(1998) who obtained 892 redshifts in the fields of nine distant
cluster candidates selected from the GHO and PDCS catalogues.  Three
of their nine candidate clusters showed no significant peak in the
observed redshift histogram, and three others showed between two and
four equally significant peaks at very different redshift,
illustrating the severity of projection effects in optically selected
cluster samples. By way of contrast, Castander et al.\ (1994)
demonstrate how X-ray observations can be used efficiently to test
whether optically selected distant clusters are indeed gravitationally
bound, massive systems. Castander and coworkers analyzed PSPC
observations of five GHO clusters which had spectroscopic redshifts
ranging from 0.7 to 0.92. They detected only two of the five, and
found all five to have measured X-ray luminosities, or upper limits,
of less than $1\times 10^{44}$ erg s$^{-1}$ ($0.5-2.0$ keV); i.e. they
are, at best, poor clusters which, unless detected in very large
numbers, do not provide stringent constraints on either the rate of
cluster evolution or the cosmological parameters of structure
formation models.

\section{The WARPS cluster survey}
\label{warps}

The goal of the Wide Angle Rosat Pointed Survey (WARPS) is to compile
a complete and unbiased, X-ray selected sample of clusters of galaxies
from serendipitous detections of X-ray sources in deep ROSAT PSPC
pointings. A comprehensive overview of the scientific goals of the
project, the X-ray source detection algorithm employed (Voronoi
Tesselation and Percolation: VTP), the sample selection and flux
corrections techniques, as well as first results, are presented in
Paper I. VTP is particularly well suited for the detection and
characterization of low-surface brightness emission (Ebeling \&
Wiedenmann 1993; Ebeling et al., 1996; Paper I) and is likely to
recognize even very distant clusters as extended X-ray sources (Paper
I). However, in order to reduce possible incompleteness in our cluster
sample due to erroneous classification of distant clusters as point
sources, our optical follow-up observations are not limited to
extended X-ray sources but also include likely point sources without
obvious optical counterparts (Paper II). Paper II also discusses the
WARPS $\log N-\log S$ distribution of poor clusters of galaxies and
its implications for cluster evolution.

The first two WARPS papers focus on results for a complete sample of
clusters compiled over a geometric solid angle of 14.7 square degrees
during the first phase of the project. In May 1997, the WARPS project
went into its second phase which increases the total solid angle to 73
deg$^2$ and will yield a statistically complete sample of more than 70
X-ray selected galaxy clusters at $z>0.3$.  In this second phase,
cluster candidates without obvious optical counterpart on the POSS
plates (as provided by the Digitized Sky Survey) were imaged at the
Michigan-Dartmouth-MIT 1.3m and University of Hawaii 2.2m telescopes
in preparation for spectroscopic follow-up observations at larger
telescopes. Although observations of a few very distant cluster
candidates have yet to be performed (scheduled for spring 2000), the
WARPS cluster sample is already complete at $z<0.84$ over the full
solid angle.

\section{ClJ0152.7--1357}
\label{disc_obs}

In 1996 and 1997, the cluster ClJ0152.7--1357 was discovered
independently in the RDCS and WARPS surveys. Later, ClJ0152.7--1357
was also detected in the Bright SHARC survey (Nichol et al., 1999). In
this section we describe the discovery of ClJ0152.7--1357 in the WARPS
survey and discuss and summarize previous and subsequent X-ray
observations of this system.

\subsection{Discovery in the WARPS survey}

The standard WARPS X-ray analysis detected ClJ0152.7--1357 as a very
extended source 14.2 arcmin off-axis in a 20\,ks PSPC pointed
observation of NGC\,720; the POSS-2 Digitized Sky Survey image is
blank at the position of the source.  Figure~\ref{iband} shows an I
band image of ClJ0152.7--1357, taken with the UH 2.2m telescope on Aug
4, 1997, with adaptively smoothed PSPC X-ray flux contours overlaid. A
blow-up of the central cluster region is shown in
Figure~\ref{iband_core}. Based on the X-ray source extent and the
observed overdensity of faint galaxies at and around the position of
the X-ray source, it was classified as a likely distant cluster of
galaxies. The X-ray emission from ClJ0152.7--1357 shows a high degree
of substructure and a pronounced elongation along a position angle of
about 40$^{\circ}$ which follows roughly the distribution of galaxies
in the cluster core (see Section~\ref{morph} for a discussion of the
dynamical state of ClJ0152.7--1357).

On Aug 11, 1997 we observed a total of 14 distant cluster candidates,
among them ClJ0152.7--1357, with the low-resolution spectrograph LRIS
(Oke et al.\ 1995) on the Keck-II 10m telescope on Mauna Kea. Using a
longslit of $1.5''$ width and the 300/5000 grating which provides 2.4
\AA/pixel resolution and spectral coverage from 5000 \AA\ to 10000
\AA, we obtained spectra of six galaxies (see Fig.~\ref{iband_core})
close to the peak of the X-ray emission from ClJ0152.7--1357 and found
redshifts as listed in Table~\ref{rstab}. The spectra are shown in
Fig.~\ref{spectra}. All redshifts are accordant and consistent with a
cluster redshift of $z=0.8325$. All spectra show absorption features
typical of old stellar populations in elliptical galaxies, and none
shows emission lines that would suggest AGN contamination.

\subsection{Other X-ray observations of ClJ0152.7--1357}

NGC\,720 was observed not only with the ROSAT PSPC in 1992, but also
with the EINSTEIN IPC in 1980, and with the ROSAT HRI in 1994.  We
examine and compare the images from all three observations in
Fig.~\ref{xcomp}. The images are shown in chronological order and have
been registered using the astrometry solutions in the respective FITS
headers.  To allow an assessment of the quality of the raw data as
well as of the presence and morphology of any extended emission, we
show contours of the smoothed emission (a Gaussian smoothing kernel
with $\sigma=30$ arcsec was used) overlaid on the observed raw photon
data. The latter are binned such that they slightly oversample the
point spread function of the respective instrument which is
represented by the FWHM bar in the lower left corner of each image.

In the following we discuss the serendipitous HRI and IPC observations
of ClJ0152.7--1357 in more detail before summarizing briefly the
results of a recent targeted observation of the cluster with the
BeppoSAX satellite.

\subsubsection{ROSAT HRI}

The relatively high angular resolution of the ROSAT HRI ($\sim 12$
arcsec at an off-axis angle of 14 arcmin) allows us to investigate
whether contaminating point sources might contribute to the observed
PSPC flux of ClJ0152.7--1357. At an off-axis angle of 14 arcmin a
point source with a flux of about one third of the flux detected from
ClJ0152.7--1357 would be detectable with the HRI at greater than
$5\sigma$ significance. However, a secure detection of diffuse
emission from ClJ0152.7--1357 with the HRI would require an exposure
time in excess of 100 ks. As can be seen in the rightmost panel of
Fig.~\ref{xcomp}, no point sources are detected within the contours
shown in Fig.~\ref{iband} but there is marginal evidence of
low-surface-brightness excess emission at the position of the cluster,
indicating that the overwhelming majority of the emission detected
with the PSPC originates from the cluster. Moreover, the emission
detected with the HRI shows a clear elongation along the same position
angle of about 40$^{\circ}$ as the one found in the PSPC data.
Although the southwestern extension of the emission detected with the
PSPC does not coincide with any prominent galaxy overdensity in the
UH2.2m image (cf.\ Fig.~\ref{iband}), we note that, if any of the
major X-ray surface brightness peaks in the PSPC image were due to a
single, unvarying point source, they would have been detected with the
HRI.

\subsubsection{EINSTEIN IPC}
\label{ipc_disc}

It is noteworthy that ours is in fact not the first detection of
ClJ0152.7--1357 at X-ray wavelengths. As mentioned before, NGC\,720
was not only observed with ROSAT but was, in 1980, also a pointing
target of observations with the EINSTEIN observatory. A source at
$\alpha=01^h\, 52^m\, 42.8^s$, $\delta = -13^{\circ}\, 57'\, 49''$
(J2000) (i.e.\ within one arcmin of the PSPC position of
ClJ0152.7--1357) is clearly detected with the Imaging Proportional
Counter (IPC, sequence number of pointing I\,5769); the EINSTEIN
source catalogue assigns this source the number 496 and quotes a
significance of detection of 4.8$\sigma$ in the IPC broad band and
within the detect cell (Harris et al., 1990).  We show the IPC broad
band data around this position in the leftmost panel of
Fig.~\ref{xcomp}. The position angle (approximately zero) of the
apparent elongation of the source is different from the one found from
the ROSAT PSPC and HRI data. However, since the IPC point spread
function has a FWHM of about 90 arcsec FWHM in the broad band, the
source elongation found in this short IPC observation is only
marginally significant.  The same IPC source is also listed as EXSS
0150.2--1411 in the catalogue of extended EINSTEIN detections compiled
by Oppenheimer, Helfand \& Gaidos (1997) who find the source
significance (presumably in the broad band) to be 4.7 and 5.7$\sigma$
within circular apertures of 1.25 and 2.35 arcmin radius. Although
this source therefore appears to be sufficiently significant to be
included in the EMSS catalogue, it remained un-identified until its
re-discovery in the RDCS and WARPS surveys in 1996/97. We will come
back to the IPC detection of ClJ0152.7--1357 in
Section~\ref{emss_0152}.

\subsubsection{BeppoSAX}
\label{beppo}

A recent pointed BeppoSAX observation of ClJ0152.7--1357 allowed the
temperature and metallicity of the intra-cluster gas to be measured:
Della Ceca et al.\ (1999) report values of k$T=6.5^{+1.7}_{-1.2}$ keV
and $Z=0.53^{+0.29}_{-0.24}$. This gas temperature is consistent both
with the temperature estimate of $5.9^{+4.4}_{-2.1}$ keV obtained by
us from the PSPC data (cf.\ Fairley et al., 1999) and with the cluster
X-ray luminosity--temperature relation as determined by Allen \&
Fabian (1998) which predicts k$T=7.8$ keV. The relatively poor angular
resolution of the BeppoSAX telescope does not allow any conclusions to
be drawn about the possibility of point source contamination.

\subsection{X-ray properties}

Using the Galactic neutral Hydrogen column density in the direction of
the cluster of $1.47 \times 10^{20}$ cm$^{-2}$ (Dickey \& Lockman
1990) as well as a metallicity of 0.5 and a gas temperature of 6.5 keV
(Della Ceca et al., 1999), we convert the total PSPC count rate of
$(0.0237 \pm 0.0015)$ ct s$^{-1}$ (PHA channels 50 to 200) measured in
the WARPS analysis into a total, unabsorbed flux of $(2.90 \pm 0.18)
\times 10^{-13}$ \ergps cm$^{-2}$ (0.5--2.0 keV), corresponding to an
X-ray luminosity of $(8.59 \pm 0.53) \,[(15.5\pm 0.95), (33.7\pm
2.08)] \times 10^{44}$ \hmtwo \ergps in the 0.5--2.0 keV [0.3--3.5
keV, bolometric] band. Thus, ClJ0152.7--1357 is slightly more luminous
than MS1054.4--0321, making it the most X-ray luminous distant cluster
detected so far. It is also worth noting that both ClJ0152.7--1357 and
MS1054.4--0321 are more X-ray luminous than any other known cluster at
$z>0.55$.

We find our measurement of the total, unabsorbed cluster flux of
ClJ0152.7--1357 in the 0.5--2.0 keV band to be in good agreement with
all other results obtained from the available X-ray observations of
this system:
\begin{description}
\item[EINSTEIN IPC:] $f(<1.5\, h_{\rm 50}^{-1}\, {\rm Mpc}) = (3.76 \pm
      0.84) \times 10^{-13}$ \ergps cm$^{-2}$ (this work)
\item[ROSAT PSPC (RDCS):] $f(<1.5\, h_{\rm 50}^{-1}\, {\rm Mpc}) = (2.2 \pm
      0.2) \times 10^{-13}$ \ergps cm$^{-2}$ (Della Ceca et al., 1999)
\item[ROSAT PSPC (WARPS):] $f({\rm total}) = (2.90 \pm
      0.18) \times 10^{-13}$ \ergps cm$^{-2}$ (this work)
\item[ROSAT PSPC (Bright SHARC):] $f({\rm total}) = (2.93 \pm
      0.16) \times 10^{-13}$ \ergps cm$^{-2}$ (Romer et al., 1999)
\item[ROSAT HRI:] $f(<1.5\, h_{\rm 50}^{-1}\, {\rm Mpc}) = (2.8 \pm
      1.1) \times 10^{-13}$ \ergps cm$^{-2}$ (this work)
\item[BeppoSAX:] $f(<2.0\, h_{\rm 50}^{-1}\, {\rm Mpc}) = (1.9 \pm
      0.4) \times 10^{-13}$ \ergps cm$^{-2}$ (Della Ceca et al., 1999)
\end{description}

All ROSAT flux measurements agree within the errors\footnote{note that
the RDCS and HRI measurements use a fixed metric aperture whereas
WARPS and Bright SHARC measure the total cluster flux}. By comparison,
the IPC result is high and the BeppoSAX result is low; compared
directly, the discrepancy between these two measurements is
significant at the $2\sigma$ confidence level.

As noted before in Section~\ref{beppo}, the measurements of the
cluster gas temperature obtained independently from the PSPC data
(this work) and the BeppoSAX data (Della Ceca et al., 1999) also agree
well within their errors.

Although we cannot rule out that some of the observed emission
originates from one or more variable point sources, the overall good
agreement of the source positions, X-ray fluxes and cluster gas
temperatures measured for ClJ0152.7--1357 between 1980 and 1998 makes
major contamination unlikely.

Table~\ref{partab} summarizes the
optical and X-ray properties of ClJ0152.7--1357.

\section{Possible systematic biases in the EMSS cluster sample}
\label{emss}

As mentioned in Section~\ref{ipc_disc}, ClJ0152.7--1357 was detected
with the EINSTEIN IPC at 4.8$\sigma$ significance (EOSCAT, Harris et
al., 1990); however, the EMSS source catalogue (Gioia et al., 1990a)
lists the respective IPC field (I\,5769) as containing no
serendipitous detections that would be significant at the greater than
4$\sigma$ level. Since this discrepancy has been the subject of some
debate, we investigate the issue in detail in the following.
Specifically, we address three questions: firstly, how can the two
catalogues, using (apparently) the same data, arrive at substantially
different significances of detection for the same source? Secondly,
what are the implications of the absence of ClJ0152.7--1357 from the
EMSS catalogue for the overall completeness of the EMSS cluster
sample? And thirdly, what are the consequences of our findings for the
clusters included in the EMSS?

\subsection{The IPC detection of ClJ0152.7--1357}
\label{emss_0152}

Both the EINSTEIN IPC source catalogue (EOSCAT, Harris et al., 1990)
and the EMSS sample (Gioia et al., 1990a) were compiled using the same
source detection algorithm. It combines a sliding cell detection
algorithm (cell geometry: $2.4\times 2.4$ arcmin$^2$) with a
maximum likelihood (ML) peak finding algorithm which fits a Gaussian
model of the instrumental point spread function (the size of which
varies with the chosen energy range) to the data inside the detect
cell. The final source positions are taken from the ML results. While
this approach is adequate for the detection of point sources, the use
of a peak finding algorithm can clearly lead to non-optimal results in
the case of extended sources with internal structure.

While the EOSCAT and EMSS results for ClJ0152.7--1357 are obtained
from the same data, the compilation procedures of the two catalogues
are not entirely identical. EOSCAT computes the source significance
within a detect cell centred on the ML source position measured in the
IPC {\em broad band} ($0.16-3.5$ keV), whereas the EMSS uses the ML
source position determined in the IPC {\em hard band} ($0.81-3.5$
keV). However, both catalogues use the {\em broad band}\/ photons
within the detect cell to compute the source significance that is used
as the final criterion for the inclusion of sources in the respective
catalogue. The rationale behind the two-band approach chosen by the
EMSS team is to take advantage of the higher resolution of the IPC in
the hard band without sacrificing the better photon statistics of the
broad band data (Maccacaro and Gioia, private communication). The
energy dependence of the instrumental resolution means, however, that
a narrower point spread function will be used by the ML algorithm in
the hard band --- which, as we shall see, is part of the reason why
the EMSS missed ClJ0152.7--1357.

\subsubsection{Re-analysis of the IPC data for ClJ0152.7--1357}

We re-analyze the IPC data for field I\,5769 in both the hard and the
broad band using the same sliding cell algorithm employed by the
EOSCAT and EMSS teams. For the broad band data our analysis yields
results similar to those listed in EOSCAT for source \#496: at the
position maximizing the source significance in the broad band we find
the detect cell to contain 44 photons of which 9.9 are expected to be
background. The resulting signal-to-noise ratio (snr) in the broad
band is 5.1. At the ML source position quoted by EOSCAT we measure 38
counts in the detect cell (EOSCAT: 44) of which 9.8 are attributed to
background (EOSCAT: 11.7); the resulting snr value is 4.6 (EOSCAT:
4.8). Our results are also in good agreement with those of
Oppenheimer, Helfand \& Gaidos (1997) who, in their independent
re-analysis of the EINSTEIN IPC data, find the significance of their
source EXSS 0150.2--1411 to be 4.7 and 5.7$\sigma$ (presumably in the
broad band) within circular apertures of 1.25 and 2.35 arcmin radius.

The left panel in Figure~\ref{emss_det} shows contours of the smoothed
X-ray emission in the IPC broad band at the position of
ClJ0152.7--1357 with both ours and the EOSCAT source position
marked. Also shown is the contour within which our analysis finds the
signal-to-noise ratio in the hard band and within the detect cell to
exceed the threshold value of four. Although the astrometry used to
create this image is taken directly from the EINSTEIN events list, the
offset of the marked EOSCAT source position from the peak of the
emission suggests that the satellite attitude solution used in the
original EOSCAT analysis may have differed by some 10 to 20
arcseconds. Note that the non-sphericity of the emission causes the
position of the X-ray peak to lie only marginally within the
$\mbox{snr}=4$ contour.

The results of the same analysis in the IPC hard band are shown in the
right panel of Figure~\ref{emss_det}. Again we show the $\mbox{snr}=4$
(broad band) contour as well as our best estimate of the source
position in the broad band. Also shown is the source position returned
by the ML algorithm from the EMSS analysis of the hard band data
(kindly provided by Isabella Gioia). Due to the more pronounced
bimodality of the source in the hard band, the ML algorithm, using a
narrower model of the point spread function than in the IPC broad
band, now centres on an apparent peak more than one arcmin north of
the position that maximizes the source significance in the broad band.
At the EMSS source position we measure a value of 4.1 for the broad
band snr; slight differences in the astrometric solution caused the
original EMSS snr measurement at this position to fall just below the
threshold value of four. Consequently, the EMSS rejected
ClJ0152.7--1357 as not sufficiently significant to be included in the
EMSS catalogue, and thus missed what would have been the most distant
and most X-ray luminous cluster in the EMSS sample.

\subsubsection{Summary of our re-analysis of the IPC data}

As demonstrated in the previous section and illustrated in
Fig.~\ref{emss_det}, the use of a peak-finding ML detection algorithm
in the EINSTEIN IPC data analysis leads to significant offsets between
the ML source position and the one maximizing the broad-band snr of
ClJ0152.7--1357. Moreover, in the presence of emission with apparent
substructure the ML source positions in different energy bands can
differ significantly. If ClJ0152.7--1357 were a spherically
symmetric, relaxed system with a radial surface brightness profile
following a beta model, the peak centering algorithm would very likely
have come closer to returning the maximal possible source significance
of $5.6\sigma$ (using the PSPC count rate and assuming a core radius
of 250 kpc). This leads us to investigate whether the failure of the
EMSS to include ClJ0152.7--1357 can be regarded as symptomatic of a
general bias against unrelaxed clusters.

\subsection{Detection bias in the EMSS cluster sample}
\label{emss_bias}

Before we attempt to assess the importance of cluster substructure for
the efficiency of the EMSS point source detection algorithm (or, more
generally, any algorithm that explicitly or implicitly assumes a
unimodal source geometry), it should be stressed that this assumption
is not vital to the source detection process. The choice of source
detection algorithm is crucial though, as the algorithm's biases can
have a significant impact on the statistical quality of the resulting
sample. The EMSS and WARPS surveys, for instance, are inherently
different due to differences in the source detection process. The EMSS
is {\em X-ray surface brightness limited}\/ (the selection criterion
is the significance of the flux in a detect cell of fixed angular
size, and the survey flux limit refers to the flux in the same detect
cell, Gioia et al., 1990a) while WARPS is almost completely {\em X-ray
flux limited}\/ (the detection procedure uses a very low surface
brightness threshold -- see Paper I -- and the limiting flux is the
total flux of the cluster including the fraction that has escaped
direct detection).

\subsubsection{Simulations of unrelaxed clusters}

We investigate the redshift and luminosity dependence of the EMSS
detection efficiency for morphologically complex sources by simulating
IPC observations of two kinds of unrelaxed clusters: firstly, mergers
of two similarly extended components (akin to ClJ0152.7--1357) and,
secondly, extended systems containing a compact but off-centre core
(similar to MS$1054-0321$, see Section~\ref{morph}).
Table~\ref{simtab} gives an overview of the model parameters used in
the simulations. In all simulations we assume a uniform background of
$2.5\times 10^{-2}$ ct s$^{-1}$ within the detect cell and an exposure
time of 2.5 ks, values typical of the average IPC pointing; we also
blur all simulated images by convolving them with a Gaussian of 33
arcsec width ($1\sigma$) thereby accounting for the IPC point spread
function (Lea \& Henry 1988). Finally we scale the total emission from
both components such that the maximal significance in the broad band
is always constant at the EMSS threshold value of $4\sigma$ within the
detect cell. Since we are investigating a systematic effect, no
Poisson noise is added to the simulated data.

Figure~\ref{sim} summarizes the results of our simulations by showing,
for a range of projected subcluster separations, the signal to noise
ratio (snr) for a detect cell centred on the overall peak of the
emission as a function of redshift.

For the merger scenario we find no evidence for a systematic
underestimation of the source significance at any redshift as long as
the projected separation of the two cluster components remains less
than about 400 kpc. This is not surprising: at redshifts greater than
0.3 such small separations are simply not resolved by the IPC.  For
projected subcluster separations of more than 400 kpc, however, the
source significance is systematically underestimated when measured
around the position of the highest peak within the emission
region. The effect is small ($\delta({\rm snr})<0.2$) but redshift
dependent.  The underestimation is most severe for X-ray luminous
systems ($L_{\rm X}> 5 \times 10^{44}$ \hmtwo \ergps, $0.3-3.5$ keV)
at intermediate to high redshift, although it takes pronounced
substructure on the scale of more than 700 kpc (in projection) to
produce a noticeable effect at $z\sim 0.8$.

For the offset-core scenario we find the redshift and luminosity
dependence to be reversed: now it is nearby clusters ($z<0.3$) of
moderate luminosity ($L_{\rm X}< 3 \times 10^{44}$ \hmtwo \ergps,
$0.3-3.5$ keV) that are most strongly affected.

These trends are consistent with the mentioned observations of
clusters at $z\sim 0.8$: while ClJ0152.7--1357 is missed by the EMSS,
MS1137.5+6625 and MS1054.4--0321 (the first apparently relaxed, the
latter a case of substructure akin to our second simulated scenario,
see Section~\ref{morph}) are both detected. Taken together, our
results thus indicate that, in the presence of different kinds of
substructure, the EMSS peak finding algorithm tends to underestimate
the significance both of nearby clusters of low to moderate X-ray
luminosity, and of distant clusters of very high X-ray luminosity.

While Figure~\ref{sim} suggests that the underestimation of the snr
within the detect cell is small (0.1--0.2), we note that the real
effect will be magnified by photon noise (not included in our
simulations) which will cause the peak position found by the EMSS ML
algorithm to deviate from the true position. The resulting positional
error is considerable: for the photon statistics of our simulated
example we find a radius of 20 arcsec for the 90\% confidence error
circle of the ML peak position. In most cases measuring the source
significance around this ML fit position will yield values that are
lower than those in Fig.~\ref{sim}. This is underlined by the very
case of ClJ0152.7--1357, a distant, X-ray luminous cluster with
substructure on the scale of 600 kpc. Its source position as
determined by the EMSS peak finding algorithm in the IPC hard band
lies so far off the X-ray centroid that the source significance in the
IPC broad band is underestimated by more than $1\sigma$ -- far more
than what is implied by Figure~\ref{sim}.

\subsubsection{Impact of the detection bias on the EMSS cluster sample}

Although the above arguments suggest that the EMSS detection bias
against unrelaxed clusters could be severe, a re-analysis of the
EINSTEIN IPC data or numerical simulations beyond the scope of this
paper would be required to accurately quantify the effect. To be
conservative, the values from our simple simulation may be taken at
face value, in which case the smallness of the amplitude of the bias
might cause one to believe that its impact on the EMSS cluster sample
will be negligible. This is, however, not necessarily true. The EMSS
catalogue as used for the definition of the EMSS cluster sample (Gioia
et al.\ 1990b) comprises 733 sources, 93 of which were identified as
clusters of galaxies. From the distribution of source significances we
estimate the number of sources with significances between 3.8 and
4$\sigma$ to be about 60; 5 of these are expected to be clusters at
redshifts greater than 0.2.  Since the fraction of significantly
unrelaxed clusters at these redshifts is almost certainly
non-negligible (see Section~\ref{morph}), and considering the inherent
uncertainties of our crude analysis, we are left with the conclusion
that the number of distant and X-ray luminous, but unrelaxed, clusters
missed by the EMSS is likely to be of the order of a few. While not
immediately alarming, this estimate is still disconcertingly high
given that the EMSS cluster sample contains only a handful of distant
X-ray luminous clusters to begin with.

\section{The X-ray morphology of distant clusters}
\label{morph}

In addition to the cosmological relevance of the sheer existence of a
distant cluster as X-ray luminous as ClJ0152.7--1357, the complex
optical and X-ray morphology of this cluster provides further
important clues. As can be seen from Fig.~\ref{iband}, ClJ0152.7--1357
consists of at least two pronounced subclusters which are (in
projection) about 600 kpc apart and are likely to merge within a few
Gyr (assuming a true spatial separation of one to a few Mpc and equal
masses of a few $10^{14} M_{\sun}$ for the two main cluster
components).

The fact that ClJ0152.7--1357 is still in the process of formation
has several interesting implications. Firstly, the subclusters
observed today are likely to have existed as separate clusters of
$L_{\rm X} \approx 4 \times 10^{44}$ erg s$^{-1}$ (0.5--2.0 keV) at a
redshift considerably greater than unity, and, secondly, the X-ray
luminosity of ClJ0152.7--1357 is bound to increase as the merger
proceeds, possibly rendering ClJ0152.7--1357 more X-ray luminous than
any cluster observed to date. Thirdly, ClJ0152.7--1357 is the third
X-ray selected cluster (out of five) detected at $z\ga 0.8$ that shows
pronounced substructure and is distinctly non-virialized, in contrast
to the morphologically much more diverse local cluster population.

The last point is illustrated by Figure~\ref{morphfig} which shows
adaptively smoothed X-ray flux contours of all three $z\sim 0.8$
clusters for which high-resolution X-ray images are currently
available.  Our HRI data reduction corrects for particle background as
well as exposure time variations using software kindly provided by
Steve Snowden. For each cluster we align and merge all available
observations which yields total exposure times as follows.
MS1137.5+6625 ($z=0.782$): 98.0 ks; MS1054.4--0321 ($z=0.829$): 186.6
ks; RX\,J1716.6+6708 ($z=0.813$): 167.2 ks. The final images use a
pixel size of $2.5\times 2.5$ arcsec$^2$ thus slightly oversampling
the HRI point-spread function.  Using {\sc Asmooth} (Ebeling et al.\
1999) the HRI counts image is then adaptively smoothed with a Gaussian
kernel the size of which is adjusted such that the local significance
of the signal within the kernel exceeds 99\%. The boxy thick contours
in Fig.~\ref{morphfig} mark the regions within which the signal is
high enough for this criterion to be met and within which all
structure apparent in the contour plots is thus significant at greater
than 99\% confidence. The dashed boxes illustrate the effect of a
placement of the EMSS detect cell on the highest peak in the emission
region. According to Fig.~\ref{morphfig}, the only relaxed cluster of
the three is MS1137.5+6625 while both MS1054.4--0321 and
RX\,J1716.6+6708 exhibit significantly nonspherical emission with
off-centre cores.

Although this high-redshift sample is still too small to allow more
quantitative conclusions, the rarity of relaxed systems is intriguing
and may indicate that we are beginning to actually observe the epoch
of formation of the majority of massive clusters.

\section{Summary and Caveat Emptor}
\label{summary}

The discovery of the X-ray luminous, unrelaxed galaxy cluster
ClJ0152.7--1357 in the WARPS cluster survey has important implications
for our understanding of the evolution of clusters as a function of
X-ray luminosity and redshift.

ClJ0152.7--1357 was previously detected in a pointed observation with
the EINSTEIN IPC; however, due to an underestimation of its
significance the source is missing from the EMSS
catalogue. ClJ0152.7--1357 would have been the most distant and most
X-ray luminous cluster in the EMSS sample.  Simulations of IPC
observations of unvirialized clusters show that the absence of
ClJ0152.7--1357 from the EMSS cluster sample may reflect a general
bias of the EMSS against unrelaxed, distant clusters. We cannot
currently quantify accurately the amplitude of such a bias; however,
conservative estimates suggest that of the order of a few X-ray
luminous clusters may have been missed at $z>0.3$.

We attempt to assess the frequency of significant substructure in
distant X-ray luminous clusters by comparing the X-ray morphology of
all such systems observed to date with the ROSAT HRI. Although the
resulting sample is small, we find tentative evidence that highly
unrelaxed systems such as ClJ0152.7--1357 may indeed be common at high
redshift.

An important implication of our findings is that quantitative
cosmological conclusions based on measurements of the abundance of
X-ray luminous, distant clusters ought to be regarded with
caution. Any comparison of cluster space densities with the
predictions of structure formation models assumes that the clusters
used satisfy the collapse criteria specified in those models (e.g.\
Press-Schechter). In the light of our morphological observations we
add a cautionary note that it is possible that many of these distant
systems do not yet meet these conditions. Clearly this would seriously
complicate the measurement of cosmological quantities using cluster
counts. However, it could also offer a new means to tackle these
questions through detailed observation and a dynamical analysis of
merger rates in statistically selected `proto-clusters'.

As far as the representative nature of current cluster samples is
concerned, the dynamical state of a cluster could complicate matters
beyond the detection bias discussed in Section~\ref{emss_bias}.
Numerical simulations by Ricker (1998) indicate that shock fronts
created in the primary collision of two merging clusters can increase
the total X-ray luminosity of the merging system by up to an order of
magnitude compared to the combined X-ray luminosity of the progenitor
clusters. While this effect is expected to be prominent only for less
than half the sound crossing time (typically a few times $10^8$ yr),
it may still, to some extent, counteract any detection bias against
merging clusters (see Section~\ref{emss_bias}) by causing such systems
to be preferentially detected in X-ray flux limited surveys.

If cluster mergers are indeed common at high redshift and the net
X-ray emission of these systems does not adequately distinguish
between formed and forming systems, we may be forced to develop much
more sophisticated models and data analysis strategies in order to
draw secure conclusions about the physical mechanisms and cosmological
implications of cluster evolution.

Deeper observations and more detailed analyses of a sizeable,
representative sample of distant X-ray luminous clusters are required
to conclusively address these issues.

\section{ClJ0152.7--1357: Outlook}

Observing time with Chandra's ACIS-I imaging spectrometer is scheduled
in Cycle 1 for a high-resolution X-ray study of ClJ0152.7--1357; the
cluster is also a GTO (Guaranteed Time Observation) target of XMM. In
combination with ongoing observations at optical and infrared
wavelengths from the ground these X-ray observations will allow
in-depth studies of the internal dynamics and mass distribution of
this system. A detailed optical study of the cluster galaxy population
with the Keck-2 telescope is underway and first results will be
presented shortly. For now we only mention that our recent multi-slit
spectroscopy observations yielded more than 20 accordant redshifts for
this system.

\begin{deluxetable}{llllll} 
\tablecolumns{5}
\tablewidth{0pc} 
\tablehead{
\colhead{galaxy} &
\colhead{R.A. (J2000)} &
\colhead{Dec (J2000)} &
\colhead{$m_I$} &
\colhead{redshift $z$} &
\colhead{features}}
\tablecaption{Positions (accurate to better than 1 arcsec), I band
              magnitudes, and redshifts of the galaxies with LRIS
              longslit spectra. The quoted redshift errors are the
              $1\sigma$ standard deviations of the values implied by
              the individual features listed in the last column. The
              redshift of galaxy C was not used in the computation of
              the cluster redshift. \label{rstab}}
\startdata 
A   & $01^h\, 52^m\, 44.9^s$ & $-13^{\circ}\, 57'\, 04''$ & $20.02 \pm 0.08$ & $0.8360 \pm 0.0004$ & Ca H\&K \\
B   & $01^h\, 52^m\, 43.7^s$ & $-13^{\circ}\, 57'\, 19''$ & $19.45 \pm 0.06$ & $0.8351 \pm 0.0002$ & Ca H\&K, G \\
C   & $01^h\, 52^m\, 43.8^s$ & $-13^{\circ}\, 57'\, 20''$ & $19.56 \pm 0.06$ & $0.8368           $ & 4000\AA\ break \\
D   & $01^h\, 52^m\, 42.9^s$ & $-13^{\circ}\, 57'\, 35''$ & $20.54 \pm 0.13$ & $0.8346 \pm 0.0007$ & Ca H\&K, G \\
E   & $01^h\, 52^m\, 39.8^s$ & $-13^{\circ}\, 58'\, 26''$ & $20.32 \pm 0.11$ & $0.8286 \pm 0.0005$ & Ca H\&K, G \\
F   & $01^h\, 52^m\, 39.6^s$ & $-13^{\circ}\, 58'\, 27''$ & $19.67 \pm 0.07$ & $0.8280 \pm 0.0004$ & Ca H\&K, G \\
\enddata
\end{deluxetable}

\begin{deluxetable}{ll} 
\tablecolumns{2}
\tablewidth{0pc} 
\tablecaption{Properties of ClJ0152.7--1357 based on the WARPS/PSPC and
BeppoSAX results. The quoted errors of the cluster count rate, flux,
and luminosity are based on the Poisson error of the detected
photons. The BeppoSAX results are taken from Della Ceca et al.\ (1999).
\label{partab}}
\tablehead{
\colhead{quantity} &
\colhead{value}}
\startdata 
redshift $z$                         & 0.8325 \\
$n_{\rm H}$                          & $1.47 \times 10^{20}$ cm$^{-2}$\\ 
PSPC exposure time                   & 19,912 s\\
detected VTP count rate (PHA 50--200)& $(0.0154\pm 0.0010)$ s$^{-1}$ \\
total count rate        (PHA 50--200)& $(0.0237\pm 0.0015)$ s$^{-1}$ \\
total unabsorbed flux (0.5--2.0 keV) & $(2.90\pm 0.18) \times 10^{-13}$ erg cm$^{-2}$
                                                                  s$^{-1}$\\
observed ICM temperature             & $5.9^{+4.3}_{-2.1}$ keV (ROSAT PSPC)\\
                                     & $6.5^{+1.7}_{-1.2}$ keV (BeppoSAX)\\
rest frame luminosity (0.5--2.0 keV) & $(8.59\pm 0.53) \times 10^{44}$ h$^{-2}_{\rm 50}$ erg s$^{-1}$\\
rest frame luminosity (0.3--3.5 keV) & $(1.55\pm 0.10) \times 10^{45}$ h$^{-2}_{\rm 50}$ erg s$^{-1}$\\
rest frame luminosity (bolometric)   & $(3.37\pm 0.21) \times 10^{45}$ h$^{-2}_{\rm 50}$ erg s$^{-1}$\\
\enddata
\end{deluxetable}

\begin{deluxetable}{llllll} 
\tablecolumns{7}
\tablewidth{0pc} 
\tablecaption{Parameters of the $\beta$ models used in the simulations
              described in Section~\protect\ref{emss_bias}. All models
              assume $\beta=2/3$ and are normalized such that the
              signal-to-noise ratio is constant at 4, which
              corresponds to a total of 27 counts within an optimally
              placed EMSS detect cell. $r_{c,1}$, $r_{c,2}$ and
              $L_1:L_2$ are the core radii and relative X-ray luminosities
              of the two components. Also listed are the ranges in
              redshift, projected metric separation $\Delta$, and total
              luminosity $L_{\rm tot}$ explored by our simulations.
              The quoted luminosities are computed in the 0.3--3.5 keV
              band assuming the IPC on-axis response matrix. See
              Figure~\protect\ref{sim} for an overview of the simulation
              results.\label{simtab}}
\tablehead{
\colhead{$r_{c,1}$ (kpc)} &
\colhead{$r_{c,2}$ (kpc)} &
\colhead{$L_1:L_2$} &
\colhead{$z$ range} &
\colhead{$\Delta$ range (kpc)} &
\colhead{$L_{\rm tot}$ range ($10^{44}$ \hmtwo \ergps)}}
\startdata 
250 & 250 & 2:3 & 0.2-1.0 & 200-1000  & 1.5-31.7 \\
100 & 350 & 1:4 & 0.2-1.0 & 100-\,400 & 1.6-27.1 \\
\enddata
\end{deluxetable}

\acknowledgements 

We thank the telescope time allocation committees of the University of
Hawai`i, UCO/Lick, the Kitt Peak National Observatory, and the
Michigan-Dartmouth-MIT observatory for their generous support of the
WARPS optical follow-up programme.

Several members of the EMSS team provided us with advice and much
useful information about details of the EMSS data processing and
source selection procedure. Their help is gratefully acknowledged.
Special thanks to Pat Henry for many fruitful discussions that led to
substantial improvements in the presentation and interpretation of our
results. Alexey Vikhlinin kindly proofread an early version of the
manuscript and made a number of helpful suggestions.  Thanks also to
Steve Snowden for kindly providing the latest version of his $\sc
cast\_hri$ package. This work has made use of data obtained through
the WWW interfaces to the GSFC/HEASARC and MPE ROSAT Public Data
Archives, as well as the STScI Digitized Sky Survey. HE acknowledges
financial support from SAO contract SV4-64008 and NASA LTSA grant NAG
5-8253. LRJ thanks the UK PPARC for financial support.

\newpage

\figcaption{I band image of ClJ0152.7--1357 taken with the UH 2.2m
            telescope on Aug 4, 1997 in sub-arcsec seeing. The total
            exposure time was 12 minutes. Overlaid are the adaptively
            smoothed X-ray flux contours in the $0.5-2.0$ keV band
            from the serendipitous PSPC observation of the
            cluster. The countours are spaced logarithmically by
            factors of 1.5; the lowest contour level lies a factor of
            3 above the background. The size of the Gaussian smoothing
            kernel is varied such that all features shown here are
            significant at the $3-4\, \sigma$ level (see Ebeling,
            White \& Rangarajan 1999 for details of the algorithm). 
            \label{iband}}

\figcaption{UH 2.2m I band image of the core of ClJ0152.7--1357
            (dashed region from Figure~\protect\ref{iband}). Galaxies
            A through F have measured redshifts of $z\approx 0.83$ and
            are thus confirmed as cluster members. \label{iband_core}}

\figcaption{Longslit spectra of galaxies A through F measured with
            LRIS on Keck-II on Aug 11, 1997 (integration time: 20 min
            for galaxies A, C, D, E, F; 10 min for galaxy B). The
            spectra have been vertically displaced to facilitate the
            comparison of their spectral features. The dashed lines
            show the locations of absorption features at the cluster
            redshift; the dotted lines mark the wavelengths of
            atmospheric absorption bands. \label{spectra}}

\figcaption{X-ray images of ClJ0152.7--1357 as obtained with the
            EINSTEIN IPC (1980, 0.3--3.5 keV), the ROSAT PSPC (1992,
            0.1--2.4 keV), and the ROSAT HRI (1994, 0.1--2.4 keV). For
            each image the bin size was set to about half the size of
            the FWHM of the point-spread function of the respective
            instrument. All images cover the same area on the sky and
            are centred on the same celestial position. Small
            differences in the absolute astrometry are probably caused
            by uncertainties in the attitude solutions of the
            observations.  The overlaid contours show the smoothed
            emission (Gaussian smoothing kernel, $\sigma=30''$) and
            are spaced logarithmically. See text for
            discussion.  \label{xcomp}}

\figcaption{EINSTEIN IPC images of ClJ0152.7--1357 in the instrument's
            broad (left) and hard band (right). Before countouring,
            the raw data were smoothed with a Gaussian kernel modelled
            after the EINSTEIN point-spread function ($\mbox{FWHM}
            \sim 90''$ in the broad band, and $\mbox{FWHM} \sim 80''$
            in the hard band).  The thick countours overlaid on both
            plots mark the region within which we find the
            signal-to-noise ratio within the detect cell and in the
            broad band to exceed four. Also shown is the broad band
            source position as listed in EOSCAT (plus sign in the left
            plot) and the hard band source position found in the EMSS
            analysis (plus sign in the right plot). The position
            maximizing the broad band snr in our own analysis is
            marked by a cross in both panels.  \label{emss_det}}

\figcaption{The signal to noise ratio measured within an IPC detect
            cell centred on the highest peak of the simulated emission
            from unrelaxed clusters as a function of redshift. The top
            panel shows the results of simulations assuming a merger
            of two similarly compact clusters for separations ranging
            from 100 to 1000 kpc. The results in the bottom panel were
            obtained assuming a very extended cluster with a compact
            core that is offset by between 100 and 400 kpc. In all
            cases the true (maximal) significance within the detect
            cell is $4\sigma$. The almost vertical thin lines connect
            loci of constant luminosity; the respective luminosity
            is given in units of $10^{44}$ \hmtwo \ergps ($0.3-3.5$
            keV). While the underestimation of the source significance
            caused by the use of the peak as an indicator of the best
            source position is less than $0.2\sigma$, the resulting
            bias is redshift dependent in the sense that both distant
            clusters of high X-ray luminosity and more nearby clusters
            of low X-ray luminosity are most strongly
            affected. Details of the simulations which produced these
            results are given in Section~\protect\ref{emss_bias}.
	    \label{sim}}

\figcaption{Adaptively smoothed HRI images of the three $z\sim 0.8$
            clusters for which high-resolution X-ray data are
            currently available. The size of the Gaussian kernel used
            in the smoothing process is determined from the
            requirement that the signal enclosed by the kernel
            represent a 99\% significant enhancement over the local
            background; within the bold contour this criterion is
            met. The contour levels are spaced logarithmically by
            factors of 1.2 with the lowest contour being 10\% above
            the background.  All images cover an area of $2\times 2$
            \hmtwo Mpc$^2$ at the redshift of the cluster; the dashed
            square marks the size of the EMSS detect cell centred on
            the brightest X-ray peak within the emission region. The
            effective exposure times at the location of the respective
            cluster are 167, 98, and 187 ks.\label{morphfig}}

\newpage

\begin{figure}
\tt Figure 1 too large to be submitted to astro-ph: please copy from 
ftp://hubble.ifa.hawaii.edu/pub/ebeling/warps/fig1.eps.gz
\end{figure}

\newpage

\begin{figure}
\tt Figure 2 too large to be submitted to astro-ph: please copy from
ftp://hubble.ifa.hawaii.edu/pub/ebeling/warps/fig2.eps.gz
\end{figure}

\newpage

\begin{figure}
\epsfxsize=\textwidth
\epsffile{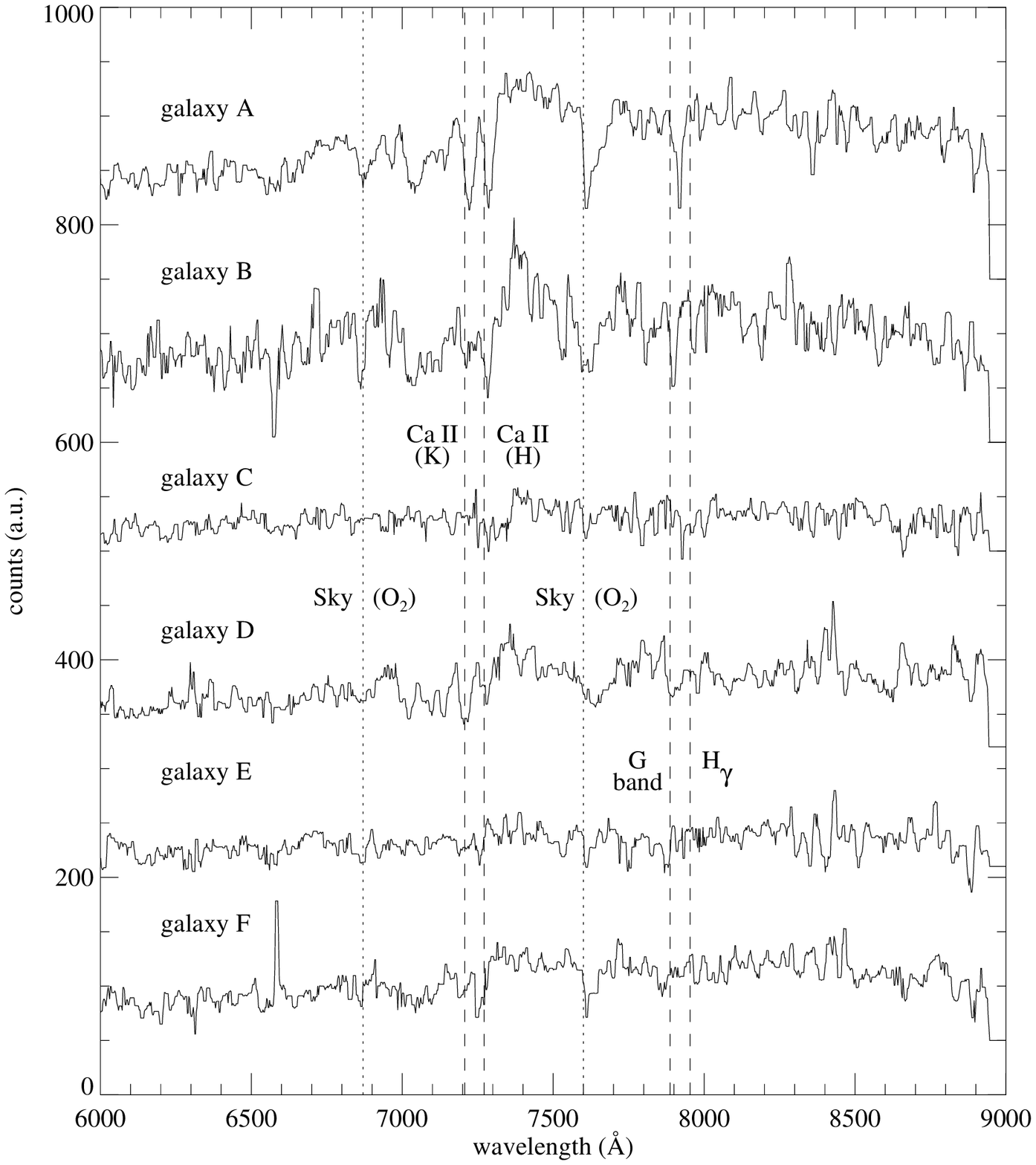}
\end{figure}

\newpage

\begin{figure}
\epsfxsize=0.45\textwidth
\centerline{\epsffile{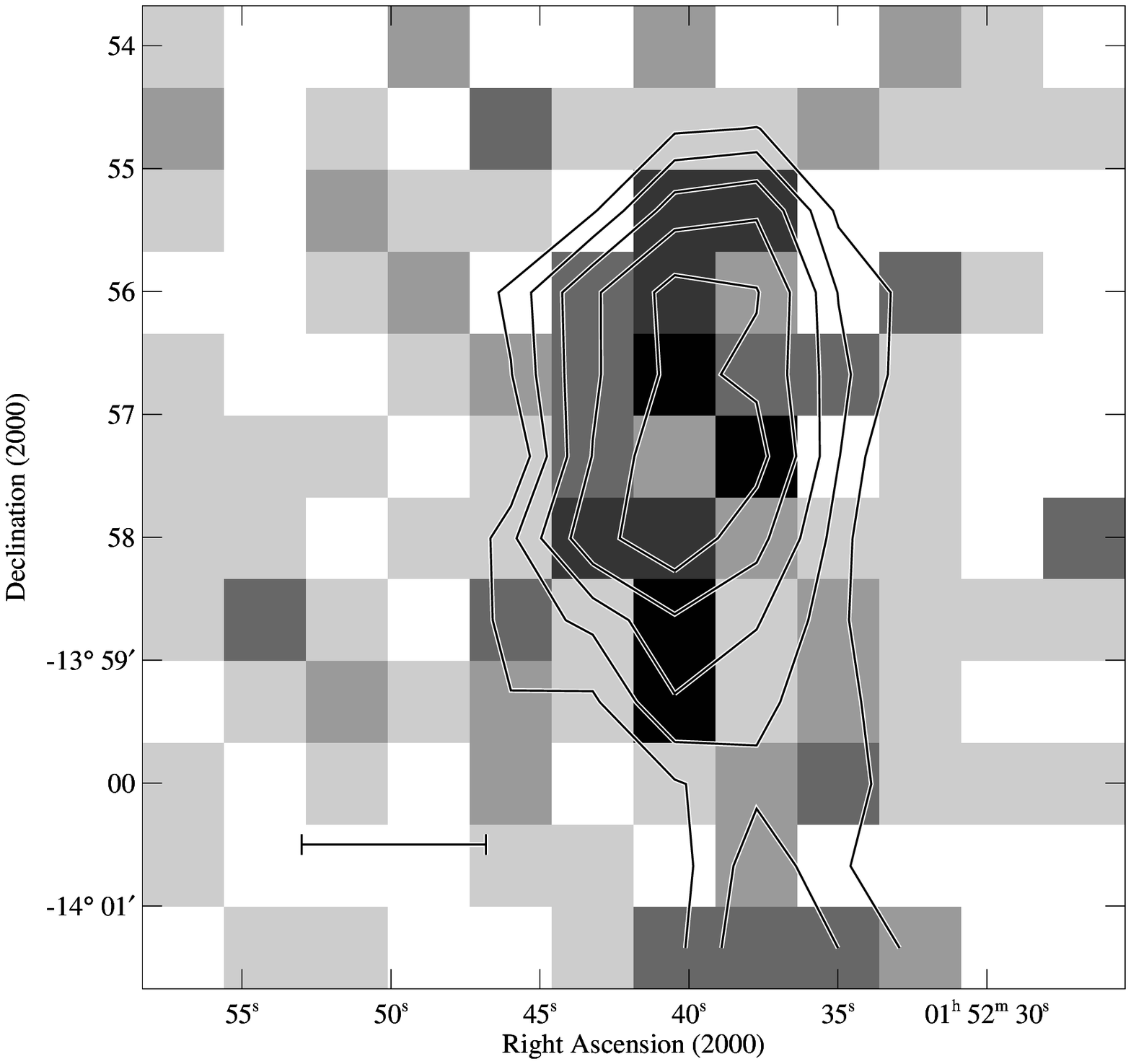}}
\epsfxsize=0.45\textwidth
\centerline{\epsffile{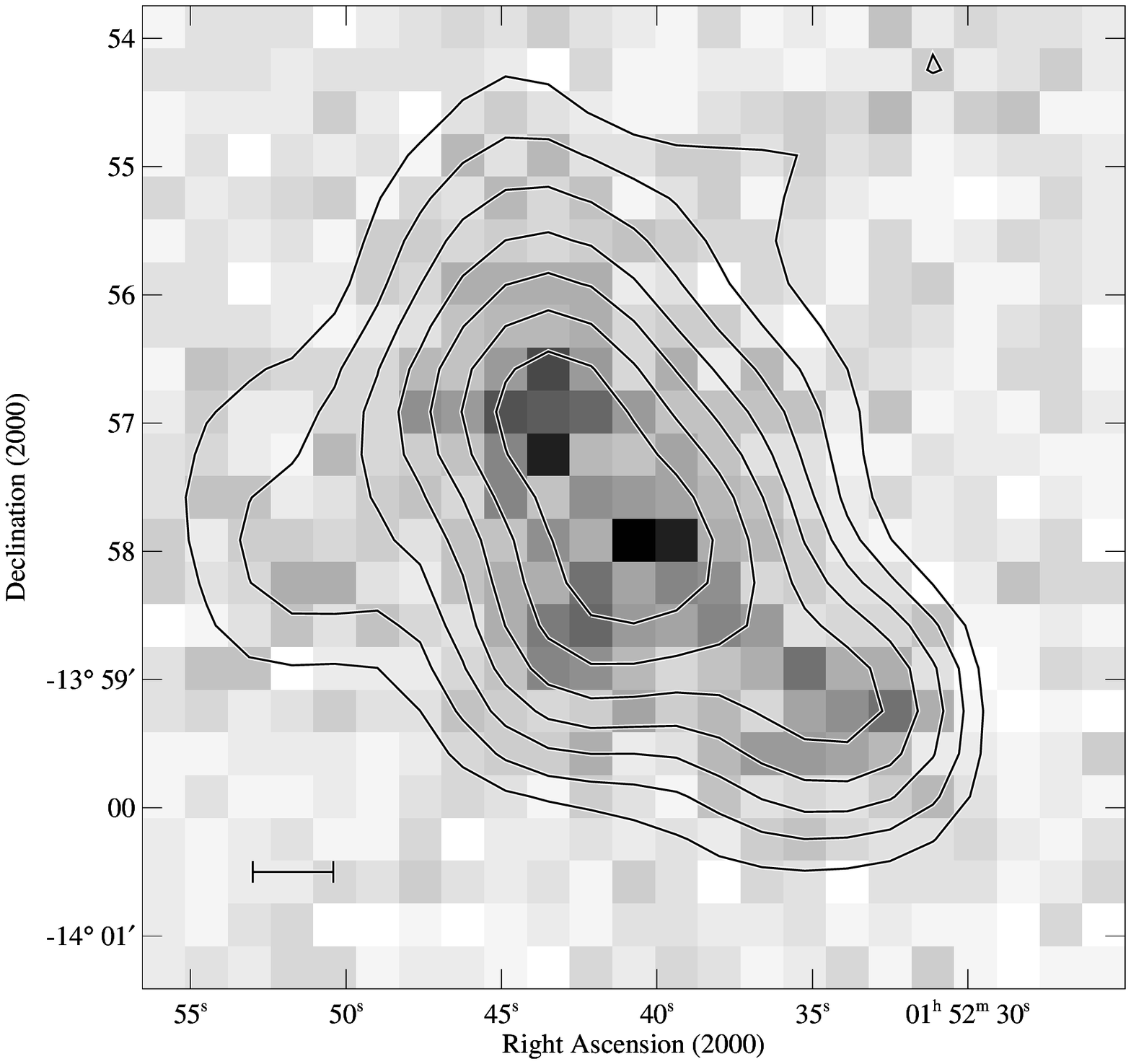}}
\epsfxsize=0.45\textwidth
\centerline{\epsffile{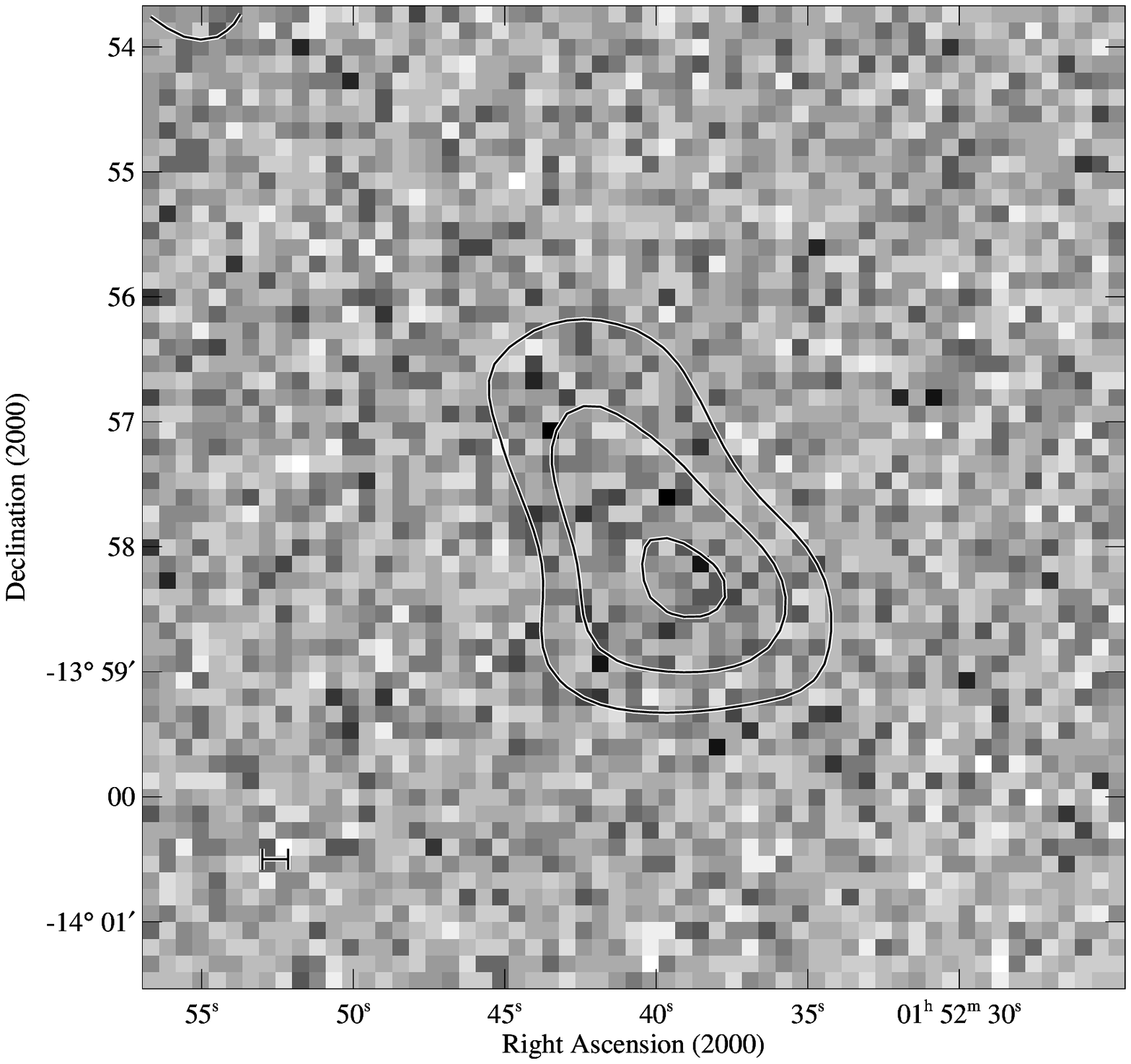}}
\end{figure}

\newpage

\begin{figure}
\epsfxsize=0.45\textwidth
\centerline{\epsffile{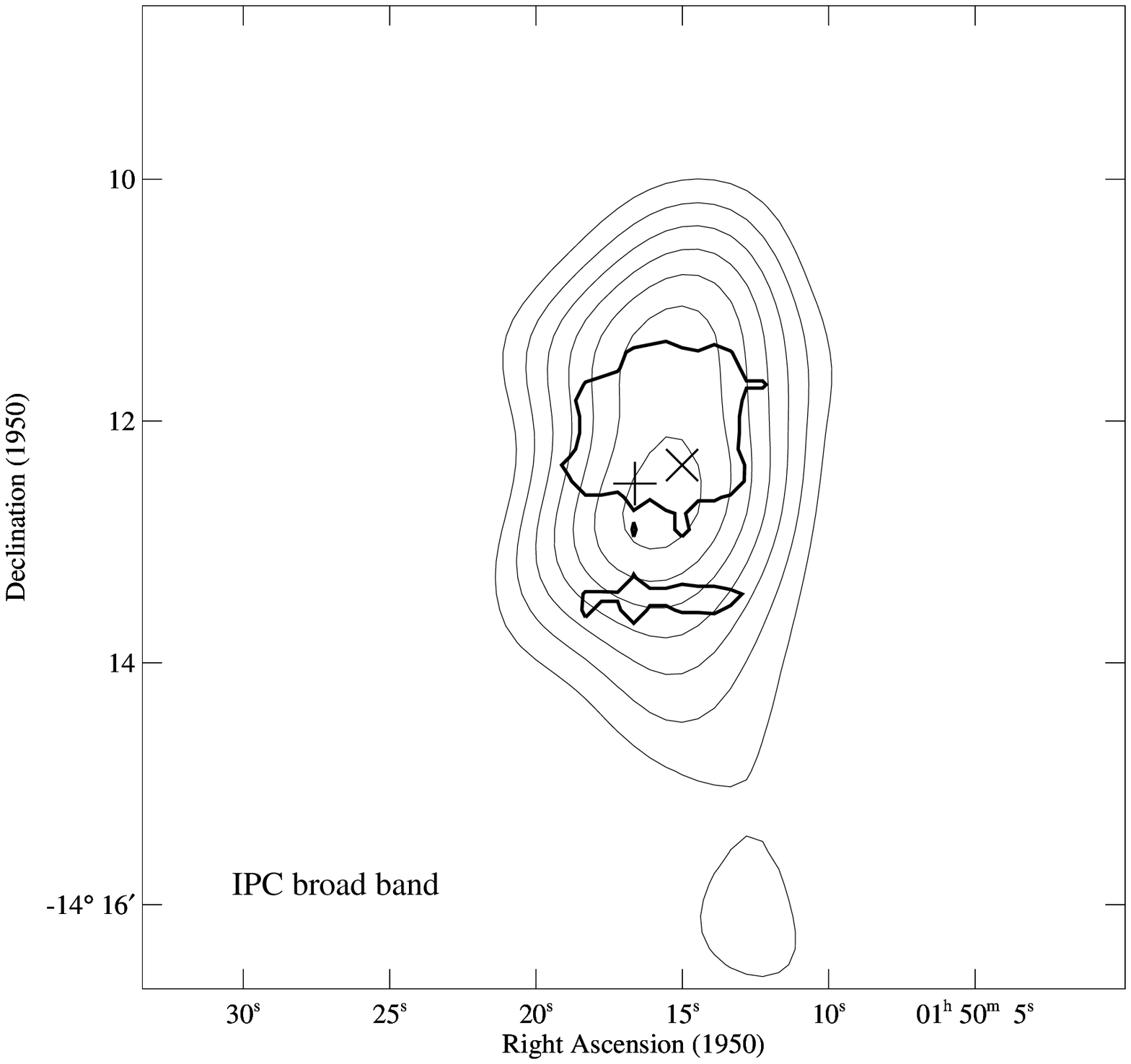}}
\epsfxsize=0.45\textwidth
\centerline{\epsffile{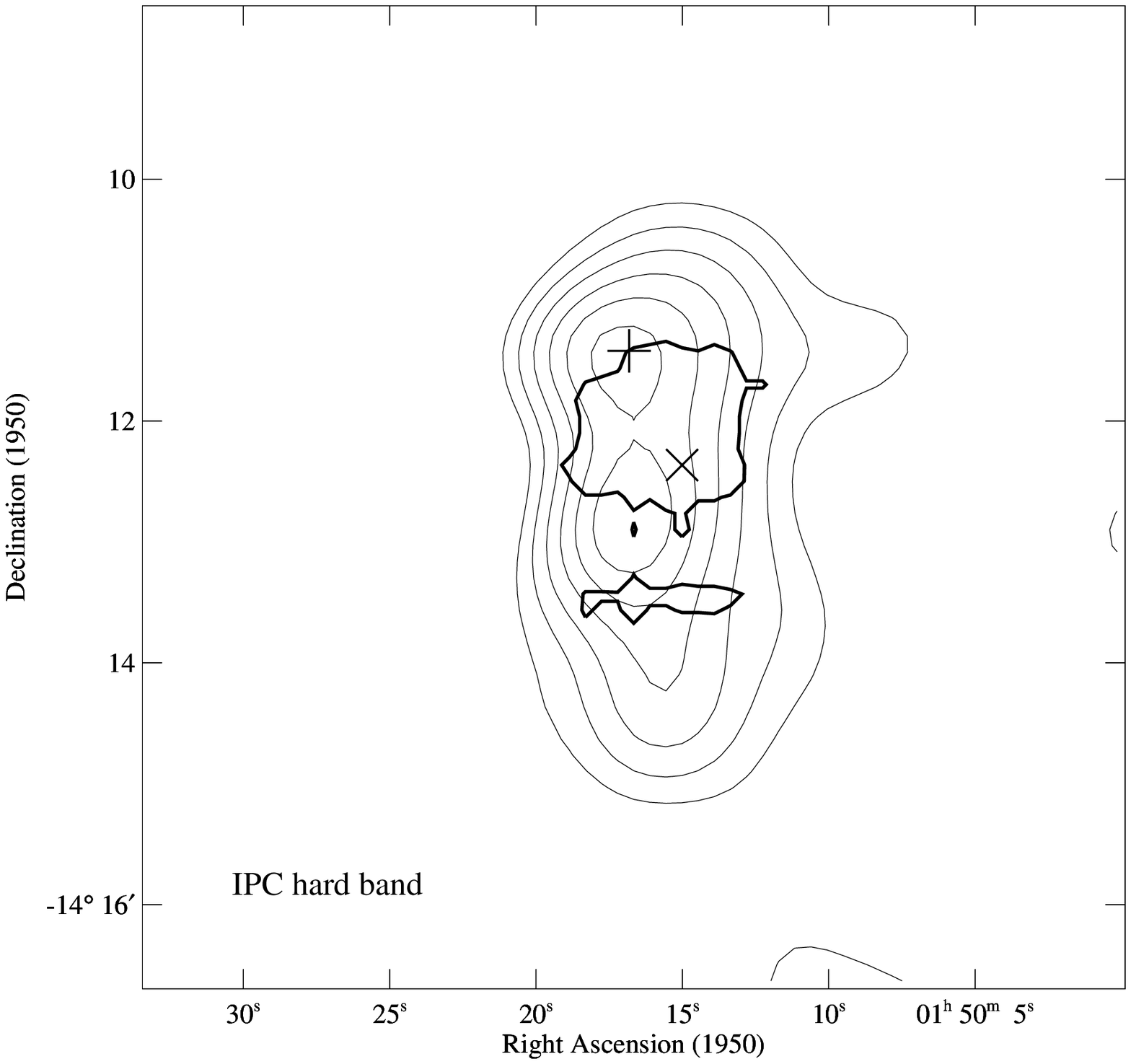}}
\end{figure}

\newpage

\begin{figure}
\epsfxsize=\textwidth
\epsffile{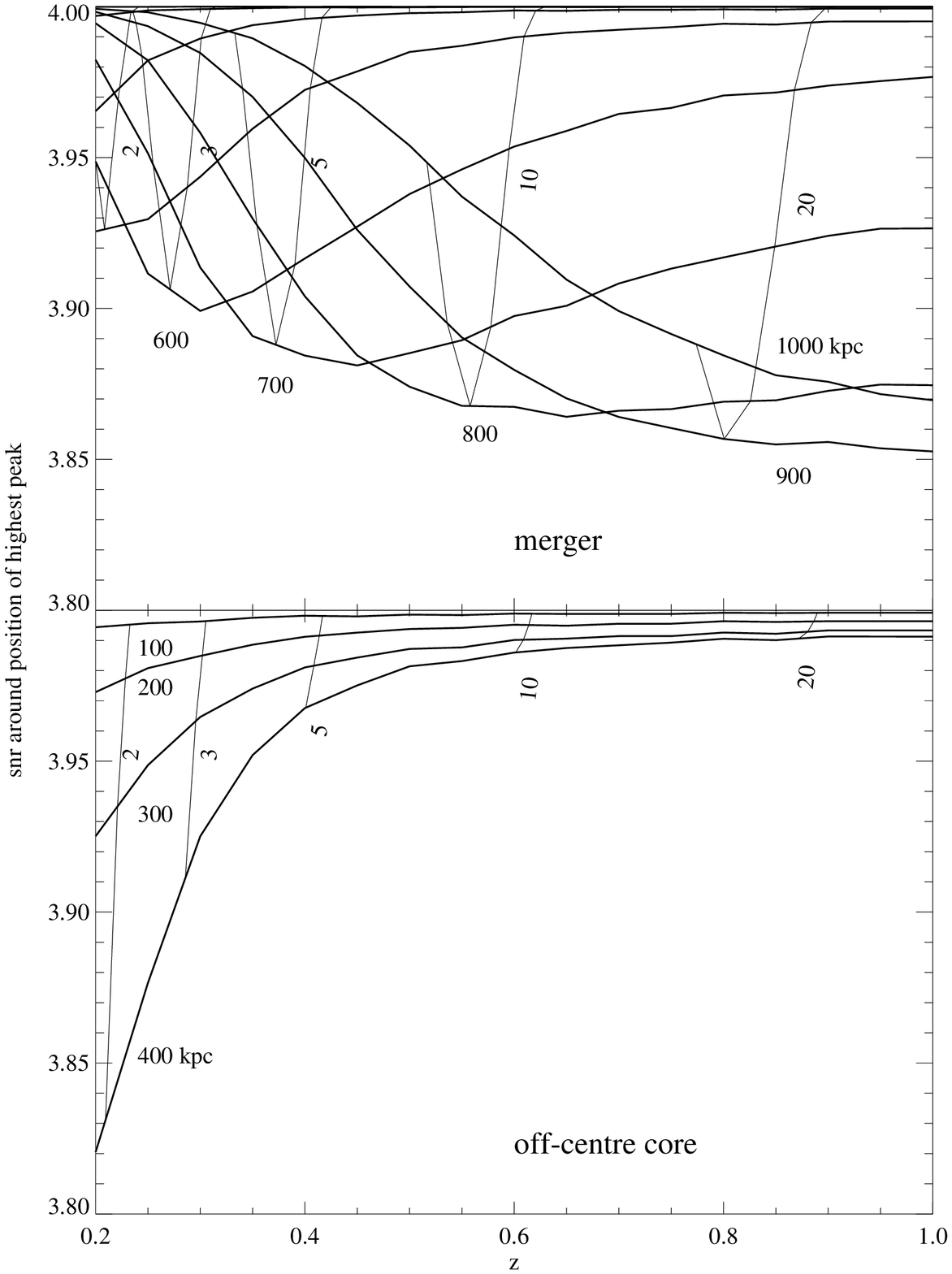}
\end{figure}

\newpage

\begin{figure}
\epsfxsize=0.45\textwidth
\centerline{\epsffile{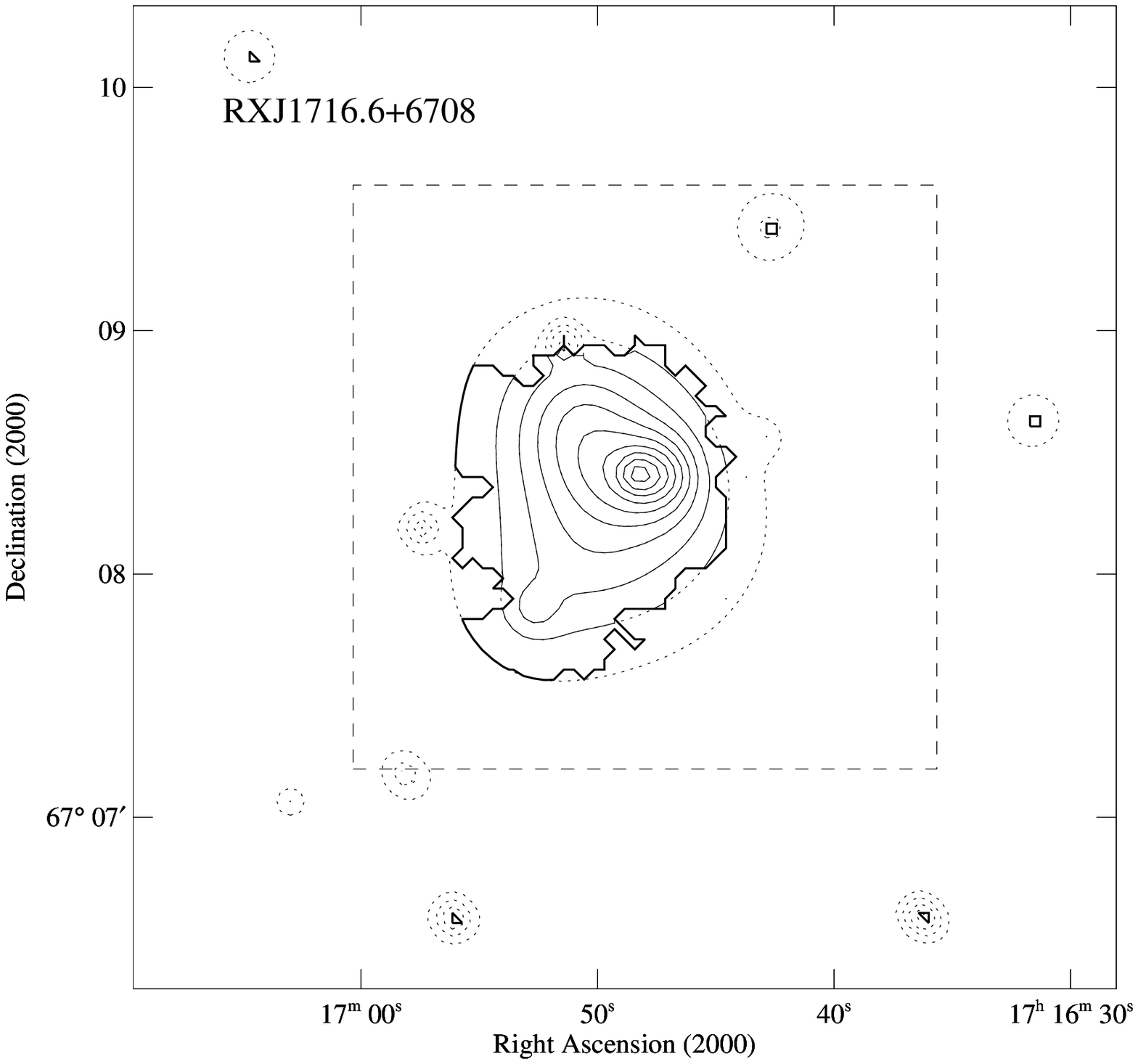}}
\epsfxsize=0.45\textwidth
\centerline{\epsffile{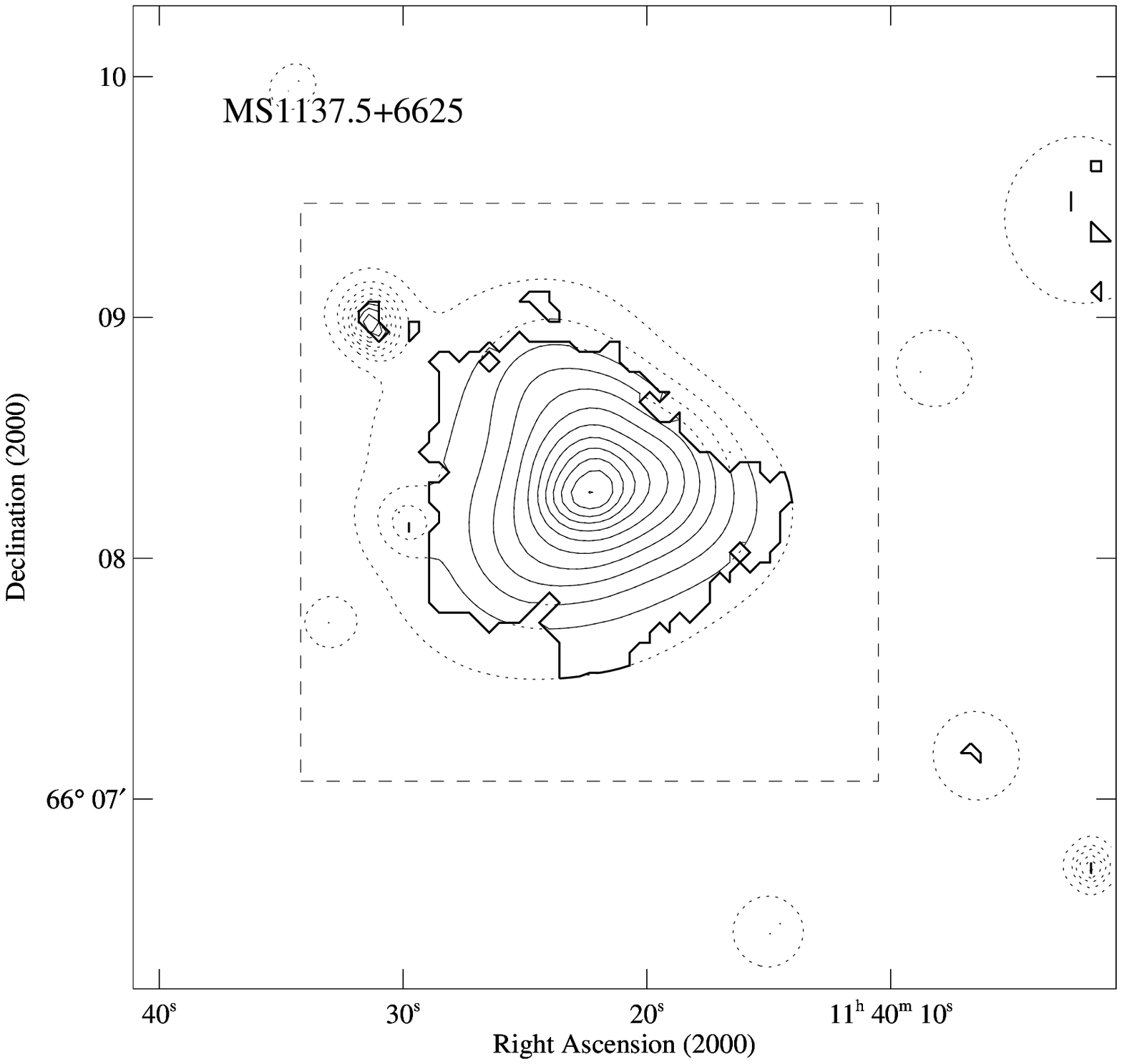}}
\epsfxsize=0.45\textwidth
\centerline{\epsffile{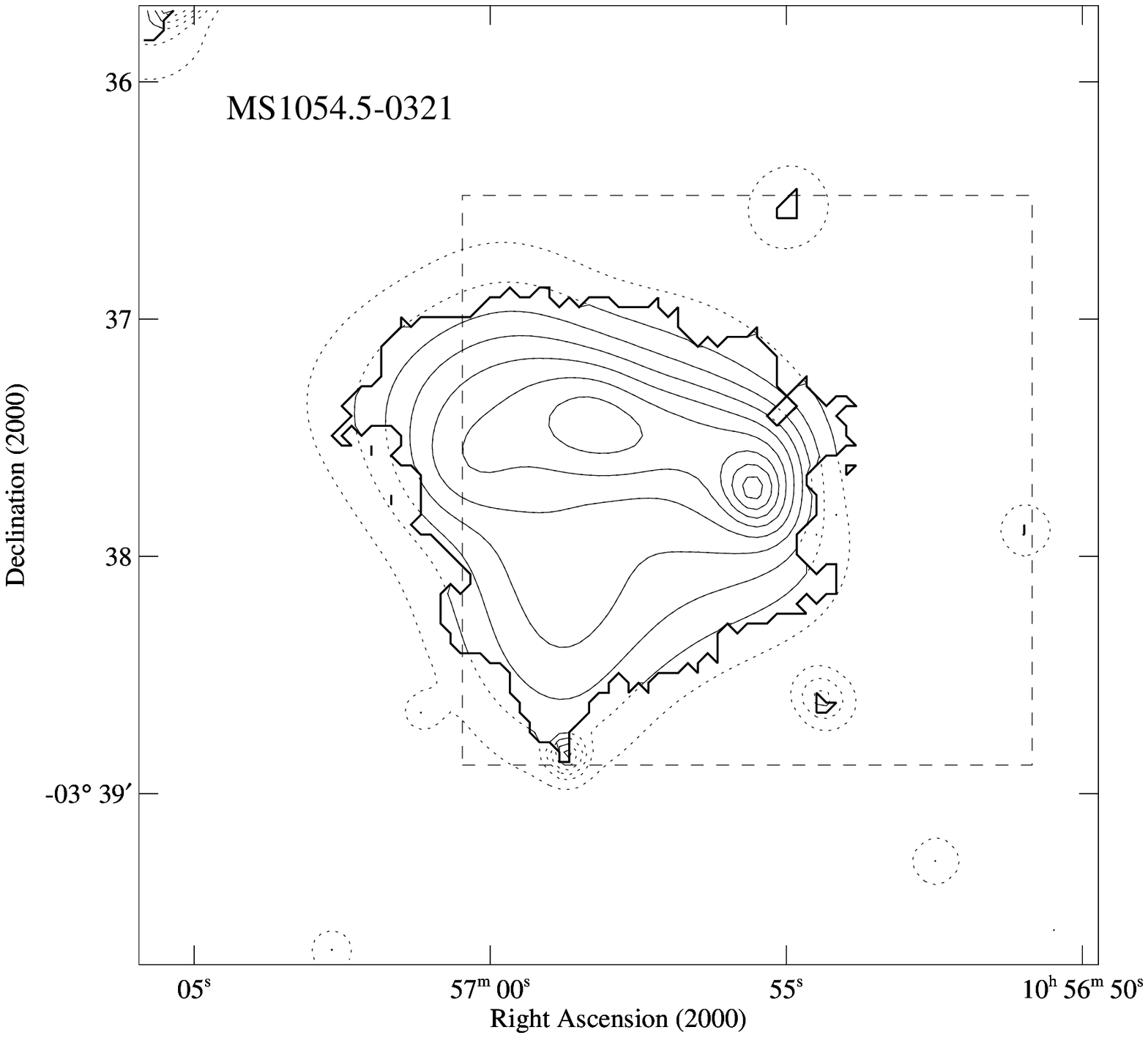}}
\end{figure}

\end{document}